\documentclass[runningheads]{llncs}
\usepackage{amsfonts}
\usepackage{amssymb}
\usepackage{graphicx}
\usepackage{mathtools}
\usepackage[utf8]{inputenc}
\usepackage[ruled]{algorithm2e}

\begin{document}

\title{Deep Reinforcement Learning for ESG financial portfolio management}
\titlerunning{Deep Reinforcement Learning for ESG financial portfolio management}
\author{Eduardo C. Garrido-Merchán, Sol Mora-Figueroa Cruz-Guzmán, María Coronado Vaca}
\authorrunning{Eduardo C. Garrido-Merchán et. al.}
\date{May 2023}

\institute{Universidad Pontificia Comillas, Madrid, Spain \\
\email{ecgarrido@icade.comillas.edu, solmora@alu.icade.comillas.edu, mcoronado@icade.comillas.edu}}

\maketitle 

\abstract{This paper investigates the application of Deep Reinforcement Learning (DRL) for Environment, Social, and Governance (ESG) financial portfolio management, with a specific focus on the potential benefits of ESG score-based market regulation. We leveraged an Advantage Actor-Critic (A2C) agent and conducted our experiments using environments encoded within the OpenAI Gym, adapted from the FinRL platform. The study includes a comparative analysis of DRL agent performance under standard Dow Jones Industrial Average (DJIA) market conditions and a scenario where returns are regulated in line with company ESG scores. In the ESG-regulated market, grants were proportionally allotted to portfolios based on their returns and ESG scores, while taxes were assigned to portfolios below the mean ESG score of the index. The results intriguingly reveal that the DRL agent within the ESG-regulated market outperforms the standard DJIA market setup. Furthermore, we considered the inclusion of ESG variables in the agent's state space, and compared this with scenarios where such data were excluded. This comparison adds to the understanding of the role of ESG factors in portfolio management decision-making. We also analyze the behaviour of the DRL agent in IBEX 35 and NASDAQ-100 indexes. Both the A2C and Proximal Policy Optimization (PPO) algorithms were applied to these additional markets, providing a broader perspective on the generalization of our findings. This work contributes to the evolving field of ESG investing, suggesting that market regulation based on ESG scoring can potentially improve DRL-based portfolio management, with significant implications for sustainable investing strategies.}


\section{Introduction}
In recent years, Deep Reinforcement Learning (DRL) has demonstrated great promise in addressing complex decision-making problems, for instance, in healthcare, DRL has been used to design personalized treatment strategies, with applications ranging from managing chronic conditions like diabetes to tailoring cancer treatment regimens \cite{komorowski2018artificial}. Within the field of autonomous vehicles, DRL has shown substantial potential for developing advanced decision-making systems, enhancing the capability of vehicles to navigate complex and unpredictable environments \cite{bansal2018chauffeurnet}. But we also find particularly interesting the application of DRL to the realm of financial portfolio management. 

Classical methods, such as the Markowitz model \cite{rubinstein2002markowitz}, have long been relied upon for determining the optimal portfolio composition under the assumptions of mean-variance optimization. However, these traditional models, while valuable, exhibit limitations in addressing the non-stationarity and high-dimensional complexity inherent in financial markets \cite{michaud1989markowitz}. DRL, with its capacity to learn optimal policies by interacting with the environment and handling large state and action spaces \cite{franccois2018introduction}, offers a compelling alternative.

The importance of Environmental, Social, and Governance (ESG) considerations in portfolio management is growing significantly due to the increasing emphasis on ethical finance \cite{henriksson2019integrating}. ESG considerations not only reflect the ethical and social responsibility of corporations but are also increasingly seen as indicators of long-term financial performance and risk mitigation \cite{giese2019foundations}. Therefore, integrating ESG factors into portfolio management is not merely an ethical choice, but also a strategic one \cite{kaiser2019risk}.

However, traditional models for portfolio management, including those that incorporate ESG factors, lack the ability to dynamically learn and adapt from sequential financial data \cite{michaud1989markowitz}. This underlines the potential for incorporating DRL in portfolio management, to dynamically integrate ESG considerations and to evolve investment strategies over time, providing an enhanced tool for ethical investing. Consequently, in this paper explores how DRL can dynamically adapt ESG information in the state space of the agent and handle markets whose regulations would include grants and taxes that are proportional to the ESG score of the portfolio.

The rest of this paper is organized as follows: The state of the art section provides a comprehensive literature review on the application of DRL in portfolio management, including studies that have incorporated ESG factors. The Proposed Methodology section presents our approach to integrating ESG factors into DRL for portfolio management and describes the DRL architectures used in our experiments. The Experiments section outlines the experimental setup and discusses the performance of our DRL agents in comparison with standard benchmarks. Finally, the Conclusions section summarizes our findings, discusses their implications, and provides directions for future research.

Through this research, we aim to advance the understanding of DRL's potential in improving ESG-based portfolio management, while furthering the incorporation of ethical considerations into modern investment strategies.

\section{State of the art}
This section seeks to present the state of the art in applying Deep Reinforcement Learning (DRL) to Environmental, Social, and Governance (ESG) financial portfolio management. The discussion is divided into three segments: classical approaches to ESG investing, quantitative AI applications to ESG, and DRL methods for portfolio management. 

ESG investing has gained significant attention in recent years due to an increased focus on sustainability, ethical corporate behavior, and regulatory standards \cite{busch2016sustainable}. The primary classical approach to ESG investing involves screening companies based on their ESG score, and these scores are often derived from third-party ESG rating agencies \cite{berg2022aggregate}. Various asset pricing models, like the Capital Asset Pricing Model (CAPM) and the Fama-French three-factor model, have been used to analyze the effect of ESG factors on portfolio returns. Moreover, optimization approaches like Markowitz’s Modern Portfolio Theory (MPT) have been applied to create optimal ESG portfolios \cite{markovitz1952portfolio}.

Despite the numerous classical approaches, quantitative AI approaches have begun to reshape the landscape of ESG investing, offering the promise of improved returns and risk management [5]. Machine learning techniques, including decision trees, support vector machines, and neural networks, have been used to predict ESG scores or company performance based on ESG factors \cite{bartram2021machine}. Recent studies have demonstrated that AI can offer unique insights into the dynamic relationships between ESG factors and financial performance, providing an edge over traditional methods \cite{d2021fundamental}. ESG portfolio investing has also been targeted using complex methodologies like Bayesian optimization, considering the ESG information and risk-performance indicators in a black-box that is modelled using a Gaussian process model \cite{garrido2023bayesian}.

Finally, regarding the approach of this work, the application of Deep Reinforcement Learning (DRL) to portfolio management offers an intriguing solution to the dynamic and complex nature of financial markets. DRL models, such as those based on Deep Q-Networks (DQN) and Actor-Critic methods, have been used to maximize portfolio returns while minimizing risk \cite{liu2021finrl}. These models leverage large-scale financial data, learn to navigate the market dynamics, and adapt their investment strategies over time \cite{liu2018practical}. The use of DRL in ESG portfolio management is an emerging field, with preliminary studies showing promising results \cite{liu2020finrl}.

\section{Proposed methodology}
We include ESG variables into the state space of the agent. We regulate the market applying grants and taxes to the return of every action that are proportional to the ESG weighted average of the portfolio. We compare the performance obtained by DRL agents without considering the market regulation and considering the regulation and show in the experiment section that we can include this regulation without incurring in a statistical loss of the portfolio performance. We begin this section by describing the portfolio management deep reinforcement learning framework, then the deep reinforcement learning algorithms that we have used and finally the methodology that has been used in this paper.

\subsection{Deep Reinforcement Learning for financial portfolio management}

First of all, we have considered to use deep reinforcement learning as we consider that the optimal ESG portfolio policy $\pi$ that we want to approximate is very complex, as it considers a continuous observation space $\mathcal{S}$ and real-valued actions $\mathcal{A}$ with complex interactions, hence being suited to approximate with a deep neural network. 

Deep Reinforcement Learning (DRL) combines elements from both Deep Learning \cite{lecun2015deep} and Reinforcement Learning \cite{sutton2018reinforcement}, demonstrating its capacity to discover optimal strategies from high-dimensional raw data inputs \cite{franccois2018introduction}. Mathematically, a DRL model is often described within the framework of a Markov Decision Process (MDP), denoted as a tuple:
\begin{align}
(\mathcal{S}, \mathcal{A}, P, R, \gamma),
\end{align}
where $\mathcal{S}$ represents the state space and $\mathcal{A}$ the action space. $P$ stands for the state transition probability matrix, $P(s'|s, a)$, which describes the probability of transitioning from state $s$ to state $s'$ given action $a$ is taken. $R(s, a, s')$ is the reward function, representing the immediate reward received after transitioning from state $s$ to $s'$ via action $a$. Lastly, $\gamma$ denotes the discount factor, a measure of the importance of future rewards. Depending on the DRL algorithm used to learn a policy $\pi(a|s)$, more hyper-parameters such as $\gamma$ are included.

The agent's behaviour is dictated by a policy $\pi(a|s)$, mapping states to actions, which specifies the probability of choosing action $a$ when in state $s$. This policy is iteratively updated to maximize the expected cumulative reward $E_{\pi}\left[\sum_{t=0}^{\infty}\gamma^tR(s_t,a_t)\right]$ where the expectation is taken over the trajectory of states and actions. In the context of finance, DRL models have been used for diverse applications, such as portfolio management \cite{liu2021finrl} or algorithmic trading \cite{liu2021finrl}, due to their effectiveness in handling complex dynamic systems under uncertain conditions.

In particular, for our work, the action space $\mathcal{A}$ considers to buy, sell or hold a percentage of every asset of the portfolio, being real-valued and of dimension dependent on the number of asset of the financial index considered by the portfolio, where an action consists of a vector $\mathbf{a}_t = [a{t,1}, a_{t,2}, ..., a_{t,T}]$ where $a_{t,j}$ is the proportion of asset $j$ in the portfolio and $T$ is the total number of assets in the portfolio.  
The state space $\mathcal{S}$ is defined as a set of relevant information at time $t$ that encapsulates market conditions, technical indicators, and ESG information of the index in consideration. Specifically, it includes Open, High, Close, Low, and Volume (OHCLV) data, a set of technical indicators and ESG ratings given by the following tuple $s\in S$:

\begin{align}
    s = [OHCLV_t, macd_t, boll_t, rsi_t, cci_t, dx_t, sma_t, ESG_t] \in \mathcal{S},
\end{align}

where MACD is the moving average convergence divergence trend-following momentum indicator, boll are the Bollinger bands given by adding and substracting the standard deviation on a simple moving average (sma, also used), rsi is the relative strength index momentum oscillator that measures the speed and change of price movements, cci is the commodity channel index momentum oscillator used to determine overbought and oversold levels and dx is the directional movement index, a momentum indicator that calculates the strength of the upward or downward trend over a given period. In our experiments, we will gain empirical evidence on the usefulness of including ESG information in a standard market and in a ESG regulated market. 

The reward $r_t$ of the DRL agent is defined as simply the daily return of the portfolio, which is the weighted average of the performance of the assets and its percentage in the portfolio. Concretely, if $\mathbf{p}_t$ denotes the daily vector of prices for each asset $j$, where $T$ is the total number of assets, and $\mathbf{w}_{j}$ is their weight in the portfolio, then the daily return $r_t$ is given by: 

\begin{align}
r_t = \sum_{j=1}^{T} w_t (p_t-p_{t-1})
\end{align}

The ESG-based DRL framework thus learns to optimize the policy based on the observed states and the associated rewards while incorporating ESG information into the decision-making process as we will describe in the following subsection.

\subsection{Deep Reinforcement Learning algorithms}
Once that we have described the framework that we have considered for ESG financial portfolio management, we now describe the technical details of the deep reinforcement learning algorithms that have been used in our experiments to learn the optimal trading policy: Proximal Policy Optimization (PPO) \cite{schulman2017proximal}, Advantage Actor Critic (A2C) \cite{babaeizadeh2016reinforcement}, Deep Deterministic Policy Gradient (DDPG) \cite{li2019robust}, Soft Actor-Critic (SAC) \cite{haarnoja2018soft} and Twin Delayed Deep Deterministic Policy Gradient (TD3) \cite{dankwa2019twin}.  

Proximal Policy Optimization (PPO) is a policy gradient method for reinforcement learning that maintains a balance between exploration and exploitation through the use of a surrogate objective function. Given a policy $\pi_{\theta}$ parameterized by $\theta$, we perform an update by optimizing the following objective function:

Let's denote $\pi_{\theta}(a|s)$ as the probability of taking action $a$ in state $s$ under policy $\pi_{\theta}$. The objective function that PPO optimizes can be simplified as:

\begin{equation}
L(\theta) = \mathbb{E}_{t}\left[\min\left(r_t(\theta) A_t, \text{clip}\left(r_t(\theta), 1-\epsilon, 1+\epsilon\right) A_t\right)\right]
\end{equation}

Here, $r_t(\theta) = \frac{\pi_{\theta}(a_t|s_t)}{\pi_{\theta_{\text{old}}}(a_t|s_t)}$ is the likelihood ratio, and $A_t$ is the advantage at time $t$, representing how much better action $a_t$ is compared to other possible actions. The clip function limits the value of $r_t(\theta)$ to the range $[1-\epsilon, 1+\epsilon]$, where $\epsilon$ is a small constant (like 0.2). The expectation $\mathbb{E}_{t}$ is taken over all time steps $t$.

The Advantage Actor-Critic (A2C) algorithm is a type of policy gradient method used in reinforcement learning. The algorithm maintains two neural network models: the Actor, which decides on the actions to take, and the Critic, which evaluates those actions. Let us denote $\pi_{\theta}(a|s)$ as the probability of taking action $a$ in state $s$ under the Actor policy $\pi_{\theta}$. The Critic estimates the value function $V^{\pi_{\theta}}(s)$ of each state $s$. The goal of A2C is to optimize the parameters of both networks so that the expected cumulative reward is maximized. This is achieved by performing an update on both networks at each step, according to the following rules: for the Actor, we use a policy gradient update, which is:

\begin{equation}
\theta_{\text{new}} = \theta_{\text{old}} + \alpha \nabla_{\theta} \log \pi_{\theta_{\text{old}}}(a_t|s_t) A^{\pi_{\theta_{\text{old}}}}(s_t, a_t)
\end{equation}

Where $A^{\pi_{\theta_{\text{old}}}}(s_t, a_t)$ is the Advantage function defined as $Q^{\pi_{\theta_{\text{old}}}}(s_t, a_t) - V^{\pi_{\theta_{\text{old}}}}(s_t)$, and $\alpha$ is the learning rate. The job of the critic is to estimate the value function as accurately as possible. To achieve this, it is updated to minimize the mean-squared error between its value estimates $V^{\pi_{\theta}}(s_t)$ and the actual returns $G_t$, resulting in the following update rule:

\begin{equation}
w_{\text{new}} = w_{\text{old}} - \beta \nabla_{w} \left(G_t - V^{\pi_{\theta}}(s_t)\right)^2,
\end{equation}

where $w$ are the parameters of the Critic, $G_t$ is the actual return, and $\beta$ is the learning rate. These updates are performed simultaneously, balancing the exploration of new policies (actor) and the exploitation of current knowledge (critic) to maximize the long-term reward.

Deep Deterministic Policy Gradient (DDPG) is an algorithm used in deep reinforcement learning that combines aspects of policy gradient methods and Q-learning. DDPG is particularly suited to environments with continuous action spaces and operates based on an Actor-Critic architecture. Let us denote $\mu_{\theta}(s)$ as the action suggested by the actor in state $s$ under policy $\mu_{\theta}$, and $Q(s, a)$ as the action-value function estimated by the Critic for action $a$ in state $s$. The Actor network in DDPG generates a deterministic policy, which maps states to actions directly. It's updated according to the policy gradient ascent rule, which is:

\begin{equation}
\theta_{\text{new}} = \theta_{\text{old}} + \alpha \nabla_{\theta} Q(s, \mu_{\theta_{\text{old}}}(s))
\end{equation}

where $\alpha$ is the learning rate. The Critic network estimates the action-value function $Q(s, a)$. It's updated by minimizing the mean squared error between its current estimate and the target value, calculated using a Bellman-like target:

\begin{equation}
w_{\text{new}} = w_{\text{old}} - \beta \nabla_{w} \left(Q(s, a) - (r + \gamma Q'(s', \mu_{\theta_{\text{old}}}(s')))\right)^2
\end{equation}

where $w$ are the parameters of the Critic, $r$ is the reward, $\gamma$ is the discount factor, $s'$ is the next state, and $Q'$ is the Critic's estimate of the next state-action value. DDPG also includes additional mechanisms like a replay buffer for storing and sampling past experiences, and target networks for stabilizing learning. This architecture and these mechanisms enable the efficient learning of optimal policies in continuous action spaces.

The Soft Actor-Critic (SAC) algorithm is a variant of the Actor-Critic methods used in  deep reinforcement learning, characterized by its incorporation of entropy regularization to encourage exploration. SAC is especially effective in environments with high-dimensional, continuous action spaces. Let us denote $\pi_{\theta}(a|s)$ as the stochastic policy parameterized by $\theta$ under which an action $a$ is taken in state $s$, and $Q(s, a)$ as the action-value function estimated by the Critic for action $a$ in state $s$. The goal of SAC is to optimize the policy to maximize the expected reward plus the entropy of the policy. The objective function for the policy update becomes:

\begin{equation}
J(\theta) = \mathbb{E}{s \sim D, a \sim \pi{\theta}} [Q(s, a) - \alpha \log \pi_{\theta}(a|s)]
\end{equation}

where $\alpha$ is the temperature parameter controlling the trade-off between exploitation and exploration, and $D$ is the experience replay buffer. The Critic network (or two critic networks in some implementations) estimates the action-value function $Q(s, a)$, which is updated by minimizing the mean squared error between its current estimate and the target value:

\begin{equation}
w_{\text{new}} = w_{\text{old}} - \beta \nabla_{w} \left(Q(s, a) - (r + \gamma (Q'(s', a') - \alpha \log \pi_{\theta_{\text{old}}}(a'|s')))\right)^2
\end{equation}

where $w$ are the parameters of the Critic, $r$ is the reward, $\gamma$ is the discount factor, and $Q'$ is the Critic's estimate of the next state-action value. By introducing the entropy term in the objective function, SAC promotes a more diverse range of actions and thus achieves a balance between exploration and exploitation.

The Twin Delayed Deep Deterministic Policy Gradient (TD3) algorithm is a model-free, off-policy reinforcement learning method designed to handle environments with continuous action spaces. It's an extension of the Deep Deterministic Policy Gradient (DDPG) algorithm, with several improvements that address DDPG's tendency to overestimate Q-values. Let us denote $\mu_{\theta}(s)$ as the action suggested by the Actor in state $s$ under policy $\mu_{\theta}$, and $Q_{1}(s, a)$ and $Q_{2}(s, a)$ as the action-value functions estimated by the two Critic networks for action $a$ in state $s$. The Actor network in TD3 is updated less frequently than the Critic networks (hence "delayed") according to the policy gradient ascent rule, which is:

\begin{equation}
\theta_{\text{new}} = \theta_{\text{old}} + \alpha \nabla_{\theta} Q_{\text{min}}(s, \mu_{\theta_{\text{old}}}(s))
\end{equation}

where $Q_{\text{min}}(s, \mu_{\theta_{\text{old}}}(s)) = \min(Q_{1}(s, \mu_{\theta_{\text{old}}}(s)), Q_{2}(s, \mu_{\theta_{\text{old}}}(s)))$ and $\alpha$ is the learning rate. The Critic networks estimate the action-value functions $Q_{1}(s, a)$ and $Q_{2}(s, a)$, and are updated by minimizing the mean squared error between their current estimates and the target values:

\begin{equation}
w_{\text{new}} = w_{\text{old}} - \beta \nabla_{w} \left(Q_{i}(s, a) - (r + \gamma Q'_{\text{min}}(s', a'))\right)^2
\end{equation}

where $w$ are the parameters of the Critic networks, $r$ is the reward, $\gamma$ is the discount factor, $Q'{\text{min}}(s', a') = \min(Q'{1}(s', a'), Q'_{2}(s', a'))$, and $Q'$ is the Critic's estimate of the next state-action value. TD3 introduces target policy smoothing, delayed policy updates, and double Q-learning to reduce overestimation bias and improve stability and performance.

\subsection{Dealing with ESG information in portfolio optimization with deep reinforcement learning}
We now provide details about the methodology that we have considered to enforce ESG financial portfolio management in the trading DRL agent. We have considered to modify the reward signal $r$ of the DRL agent to take into account the ESG performance of the portfolio and not only the financial returns. In order to do so, we perform two actions.

The first one is that we introduce into the state space of the DRL agent $\mathcal{S}$ three variables that represent the environmental, social and governance information of every asset that is included in the portfolio. We also compute a ESG indicator as the mean of those variables. Then, having included all the ESG variables in the agent, we compute an weighted mean of the ESG value in the portfolio, that we denote as $\varphi$. Let $\mathbf{w}_t$ be the vector of percentages of every asset, from a total of $A$ assets, in the portfolio at time $t$ and let $\mathbf{\epsilon}_t$ be the ESG score vector of every asset at time $t$, then:
\begin{align}
\varphi = \sum_{i=1}^{A} w_{ti} \epsilon_{ti},
\end{align}
where $\varphi$ can be interpreted as the ESG value of the portfolio, we leave to the practitioner to also use the median value or the winsorized mean as a indicator $\varphi$ for the ESG value of the portfolio, any statistical position measure could be substituted with respect to our choice, being this choice interpretable as a hyper-parameter of our model. Having computed $\varphi$, we now compute $\varphi$ assuming that every value of the weights $w_{ti}$ is $1/A$, where $A$ is the total number of assets and denote this quantity as $\psi$, that represents the ESG value of the index, or set of assets $\mathcal{A}$ where the portfolio is going to be placed. In particular, $\psi$ represents the ESG value of the diversified or market strategy.

We can now face two different scenarios, for every time $t$, the ESG value of our portfolio will be higher than the one of the diversified strategy $\varphi>\psi$ or lower or equal $\varphi<\psi$. In the reward function of the agent, we modify the reward signal by adding a grant if $\varphi>\psi$, hence making the reward higher, or a tax if $\varphi<\psi$, hence making the reward lower. Modifying the reward of the DRL agent codifies the goal, or objective function, to optimize by the DRL algorithm, hence representing our desire to obtain a ESG policy using these modifications in the reward function. We codify the ESG quality of the portfolio by making this grant or tax proportional to how much differs the ESG quality of the portfolio with respect to the index as a linear function. Let $r_t$ be the portfolio return with respect to time $t-1$, where we consider daily operations, and let $\lambda \in \mathbb{R}$ be a scalar representing a hyper-parameter that codifies the importance given to the ESG grant by the investor. In the case of the grant, the ESG portfolio return $\mathcal{R}_t$ would be equal to:
\begin{align}
\mathcal{R}_t = r_t + \lambda|r_t|(\varphi-\psi)/(10.0-\psi),
\end{align}
where we leave as a hyper-parameter of our model to modify this quantity and make it logarithmic or non-linear, considering any basis function or kernel $\phi(\cdot)$ depending on the investor needs $\phi(\varphi,\psi)$, hence making this approach more flexible to potential regulations and $10.0$ is the maximum ESG value as $\varphi \in [0,10]$. Analogously, in the case of the tax, we compute $\mathcal{R}_t$ as:
\begin{align}
\mathcal{R}_t = r_t - \lambda|r_t|(\psi-\varphi)/(\psi).
\end{align}
By performing these operations in the new reward signal $\mathcal{R}_t$ of the DRL agent we codify the importance $\lambda$ that the investor gives to the ESG information of every asset in the considered index or set of assets where the portfolio is going to be placed, being now the robotrader configured to find an ESG policy and not only a policy that maximizes the performance of the portfolio. 

\section{Experiments}
We carried out several experiments in the DJIA, IBEX35, NASDAQ 100 indexes to gain empirical evidence on the hypothesis that adding ESG information does not harm the performance of the agent and, in some cases, it even boost the performance of the DRL agent. We have already described in Section 3 the state space, action space and reward of the agent. For the agent evaluation we consider several metrics like the anual and cumulated return from a DRL perspective or the Sharpe, Calmar and Sortino ratio from a performance-risk tradeoff financial perspective. We also report other financial information of the portfolio like the annual volatility, the stability, the max drawdown indicator or the dialy value at risk. We will comment the following experiments according to that results. The training data period for all the DRL agents goes from 2009-01-01 to 2020-07-01 and the trading period goes from 2020-07-01 to 2021-10-31. Moreover, we have gained access to monthly Bloomberg ESG information about different assets. As the data is incomplete, we approximate missing ESG values from the existing ones that are closer to the date, incurring in bias that must be targeted in future work. Finally, we use the Yahoo Finance API to download the dialy financial data from 2009-01-01 to 2021-10-31.

\subsection{Toy experiment}
We first just include the market regulation to test whether a DRL agent adapts to the new proposed market without variables. Our hypothesis is that the performance of the agent must be similar, or close, to the one that it displays without the ESG variables, adapting its policy to achieve the gains of the regulated market. We just want to show that this is potentially possible, so we just compute a point estimate of the different metrics to illustrate the experiment. We use the DJIA index and the A2C algorithm for this illustrative experiments, showing several risk performance metrics of the agent that works in the regulated market and the one that works in the free market. We use 50000 timesteps, a learning rate of $\eta=2e^{-4}$ and an entropy coefficient of $c=5e^{-3}$. As we describe in the further work section, we leave hyper-parameter tuning for future research. 

\begin{figure}[h]
    \centering
    \includegraphics[width=\textwidth]{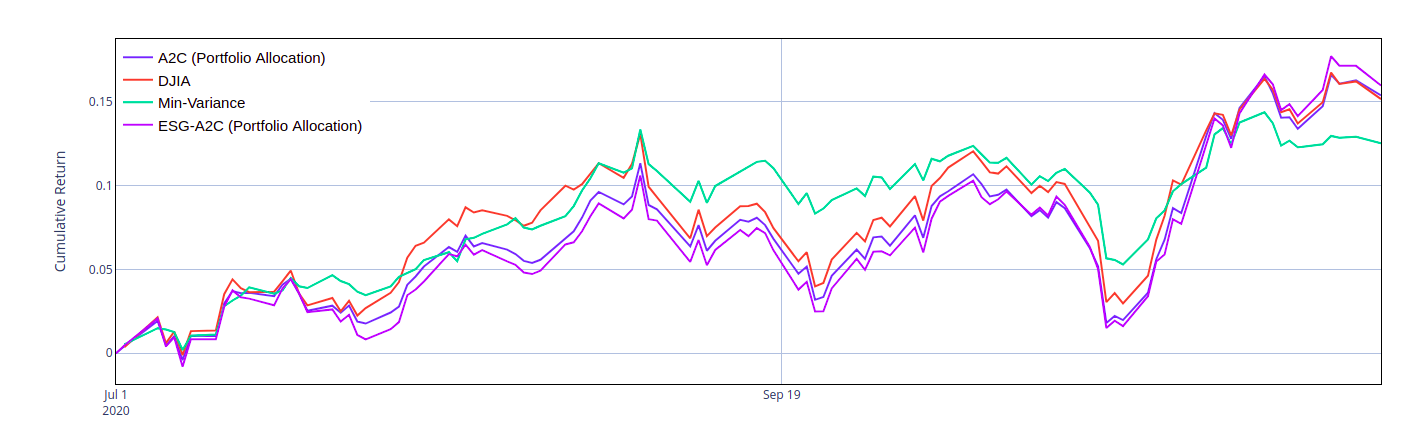}
    \caption{Performance measured with the cumulative return metric of the DRL A2C agents with respect to a min-variance baseline and the stratified index strategy (DJIA) in the trading period.}
    \label{toy_experiment}
\end{figure}

As we see in Figure $\ref{toy_experiment}$, the A2C agent that operates in the ESG regulated market is able to outperform the A2C agent in the unregulated market in terms of the cumulative return. As baselines, we have also tested the performance of a min-variance strategy and a stratified strategy, whose performance is lower than the performance of the DRL agent. We have also retrieved other measures of risk and performance that we display in Table \ref{toy_results}. 

\begin{table}[h]
    \centering
    \begin{tabular}{|c|c|c|c|}
        \hline
        \textbf{Metric name} & \textbf{DRL A2C Agent ESG Regulated} & \textbf{DRL A2C Agent} & \textbf{DJIA Baseline} \\
        \hline
        Annual return & \textbf{0.462587} & 0.435209 &  0.434549\\
        \hline
        Cumulative returns & \textbf{0.173427} & 0.164137 & 0.162246\\
        \hline
        Annual volatility & 0.161122 & \textbf{0.159924} & 0.178863\\
        \hline
        Sharpe ratio & \textbf{2.441398} & 2.340178 & 2.127074\\
        \hline
        Calmar ratio & \textbf{6.438659} & 5.949017 & 4.865749\\
        \hline
        Stability & \textbf{0.747153} & 0.740909 & 0.551107\\
        \hline
        Max drawdown & \textbf{-0.071845} & -0.073156 & -0.089308\\
        \hline
        Omega ratio & \textbf{1.482629} & 1.459303 & 1.404285\\
        \hline
        Sortino ratio & \textbf{3.733131} & 3.552500 & 3.173748\\
        \hline
        Daily value at risk & -0.018739 & \textbf{-0.018663} & -0.021025\\
        \hline 
    \end{tabular}
    \caption{Risk-performance metrics of the toy experiment in the DJIA index in the trading period according to their predictions with respect to its observable space.}
    \label{toy_results}
\end{table}

The results shown on Table \ref{toy_results} show how a regulated market does not necessarily imply a lower performance of a DRL agent and how DRL can potentially beat the market if the conditions of the trading period are similar to the ones of the training period and the training phase of the DRL is stable. In terms of performance, the DRL agent of the regulated market outperforms the other agent in terms of annual return and cumulative returns. We show how, as the market is regulated with grants and taxes, the annual volatility of the DRL agent operating in the regulated market is higher, which is logical, as the variability of the returns distribution is higher, as they are transformed with the taxes and grants. Consequently, the daily value at risk is also higher in the case of the DRL agent operating in the regulated market. However, mixed measures of risk and performance show how this risk is only slightly higher in comparison with the performance, as the Sharpe, Calmar, Omega and Sortino ratio are better in the case of the DRL agent operating in the regulated market than in the case of the DRL agent of the free market. Finally, curiously, the minimum max drawdown is also higher in the DRL regulated agent. Thus, in order to gain more robust results, we execute more experiments in the following section. 

\subsection{Benchmark experiments}
We now include ESG variables to the DRL agent operating in the regulated market and test their effect with respect to the agent without the ESG variables in its state space operated in the market without regulation. For this experiments we run several times the different DRL algorithms: A2C, PPO, DDPG, SAC and TD3; in the DJIA, IBEX-35 and NASDAQ-100 indexes, to obtain a wide variety of results that we hope that can generalize to other indexes. First, we describe the behaviour of all the algorithms in the two different markets with two different observable spaces. Then, we show a summmry of all the results, where the quantities have been standardized to get rid of ranges. 

\begin{figure}[h]
    \centering
    \includegraphics[width = 0.7\textwidth]{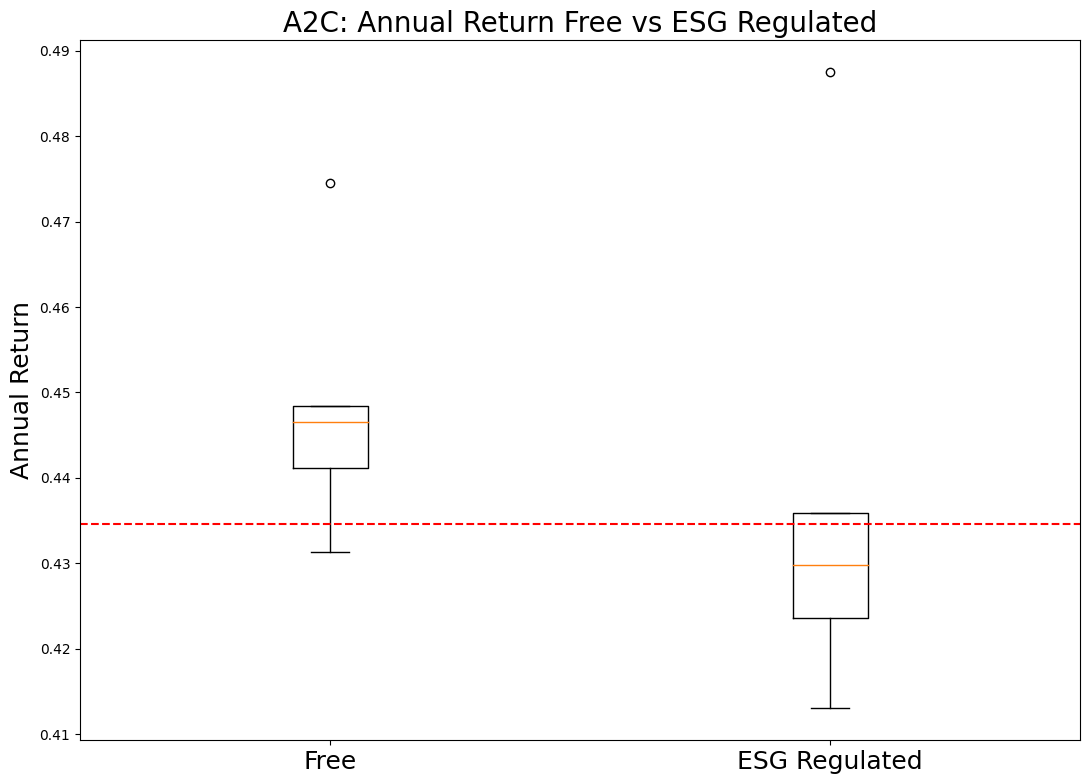}
    \caption{Box plots showing the performance measured with the annual return metric of the DRL A2C agents with respect to a stratified DJIA index strategy (red line) in the trading period.}
    \label{djia_box}
\end{figure}

We begin with the experiments of the DJIA index, where we run the $5$ algorithms with different random seeds to provide interval estimations of its behaviour. In particular, we provide the mean and standard deviation of the performance of the algorithms in both markets for the A2C algorithm in Table \ref{djia} and Figure \ref{djia_box}.  

\begin{table}[h]
    \centering
    \begin{tabular}{|c|c|c|}
        \hline
        \textbf{Metric name} & \textbf{DRL A2C Agent ESG Regulated} & \textbf{DRL A2C Agent} \\
        \hline
        Annual return & $0.4313 \pm 0.026$ & $\mathbf{0.4483 \pm 0.014}$ \\
        \hline 
        Cumulative returns & $0.1650 \pm 0.009$ & $\mathbf{0.1686 \pm 0.005}$ \\
        \hline
        Annual volatility & $\mathbf{0.1740 \pm 0.001}$ & $0.1812 \pm 0.009$ \\
        \hline
        Sharpe Ratio & $\mathbf{2.174 \pm 0.106}$ & $2.138 \pm 0.075$ \\
        \hline
        Max Drawdown & $\mathbf{-0.080 \pm 0.001}$ & $-0.083 \pm 0.004$ \\
        \hline
        Sortino ratio & $\mathbf{3.345 \pm 0.192}$ & $3.326 \pm 0.097$ \\
        \hline
        Daily value at risk & $\mathbf{-0.020 \pm 0.0001}$ & $-0.021 \pm 0.001$ \\
        \hline
    \end{tabular}
    \caption{Risk-performance metrics of the toy experiment in the DJIA index in the trading period according to their predictions with respect to its observable space for the A2C algorithm.}
    \label{djia}
\end{table}

We can see in Figure \ref{djia_box} and Table \ref{djia} how, in this case, the ESG information of the state space does not provide a stable advantage with respect to the DRL agent operating in the free market. However, due to the high volatility of the process, the maximum annual return belongs to a ESG strategy on the ESG regulated market. We can also see how lots of the experiments outperform the annual return baseline, which is plotted on red on Figure \ref{djia_box}.

Additionally, we can see how, in terms of metrics that balance risk and performance, trading with an ESG portfolio improves the quality of the portfolio in these terms according to the Sharpe and Sortino ratios and to the daily value at risk. Also, we have less volatility in the case of the ESG portfolio. 

Now we show the results of applying another DRL algorithm: proximal policy optimization (PPO). We have considered for the experiments $50000$ timesteps, a learning rate of $\eta=0.0001$, an entropy coefficient of $c=0.005$ and a batch size of $128$, leaving for further work the hyper-parameter tuning process, mainly due to its computational cost. We summarize the results in table \ref{djia_ppo}, where we can observe how, by using this algorithm, both policies are almost tied, delivering more or less the same results, again showing empirical evidence that supports the claim that using ESG portfolios does not incur in losing performance. Moreover, we can see how the ESG portfolio has slightly less volatility than the other one. 

\begin{table}[h]
    \centering
    \begin{tabular}{|c|c|c|}
        \hline
        \textbf{Metric name} & \textbf{DRL A2C Agent ESG Regulated} & \textbf{DRL A2C Agent} \\
        \hline
        Annual return & $0.4279 \pm 0.0004$ & $\mathbf{0.4286 \pm 0.0003}$ \\
        \hline 
        Cumulative returns & $0.1616 \pm 0.0001$ & $\mathbf{0.1618 \pm 0.0001}$ \\
        \hline
        Annual volatility & $\mathbf{0.1771 \pm 5.32e^{-5}}$ & $0.1772 \pm  4.773e^{-5}$ \\
        \hline
        Sharpe Ratio & $2.1000 \pm 0.002$ & $\mathbf{ 2.102 \pm  0.002}$ \\
        \hline
        Max Drawdown & $-0.0837 \pm 5.391e^{-5}$ & $\mathbf{-0.0836 \pm 2.793e^{-5}}$ \\
        \hline
        Sortino ratio & $\mathbf{3.2328 \pm 0.0031}$ & $\mathbf{3.238 \pm 0.0030}$ \\
        \hline
        Daily value at risk & $\mathbf{-0.021 \pm 7.499e^{-6}}$ & $\mathbf{-0.021 \pm 6.530e^{-6}}$ \\
        \hline
    \end{tabular}
    \caption{Risk-performance metrics of the toy experiment in the DJIA index in the trading period according to their predictions with respect to its observable space for the PPO algorithm.}
    \label{djia_ppo}
\end{table}

\begin{figure}[h]
    \centering
    \includegraphics[width = 0.7\textwidth]{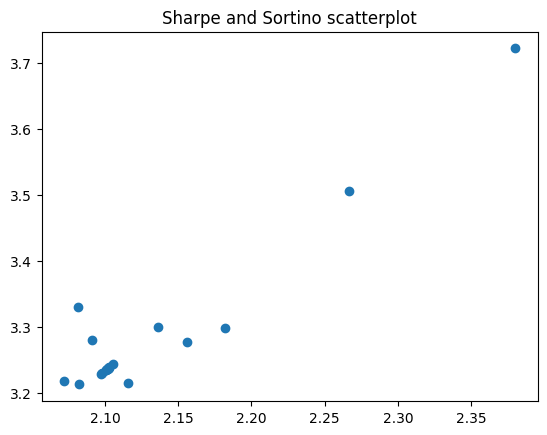}
    \caption{Scatterplot showing the direct correlation between Sharpe and Sortino values of different experiments.}
    \label{sharpe_sortino}
\end{figure}

From now on, we will report a summary of risk and performance indicators, as they exist heavy correlations between the metrics. For example, we report several Sharpe and Sortino values in the scatter plot of Figure \ref{sharpe_sortino}, where we can observe the strong correlation. Consequently, we will only report Sharpe in the next experiments, and the same happens with other metrics. 

\begin{figure}[h]
    \centering
    \includegraphics[width = 0.7\textwidth]{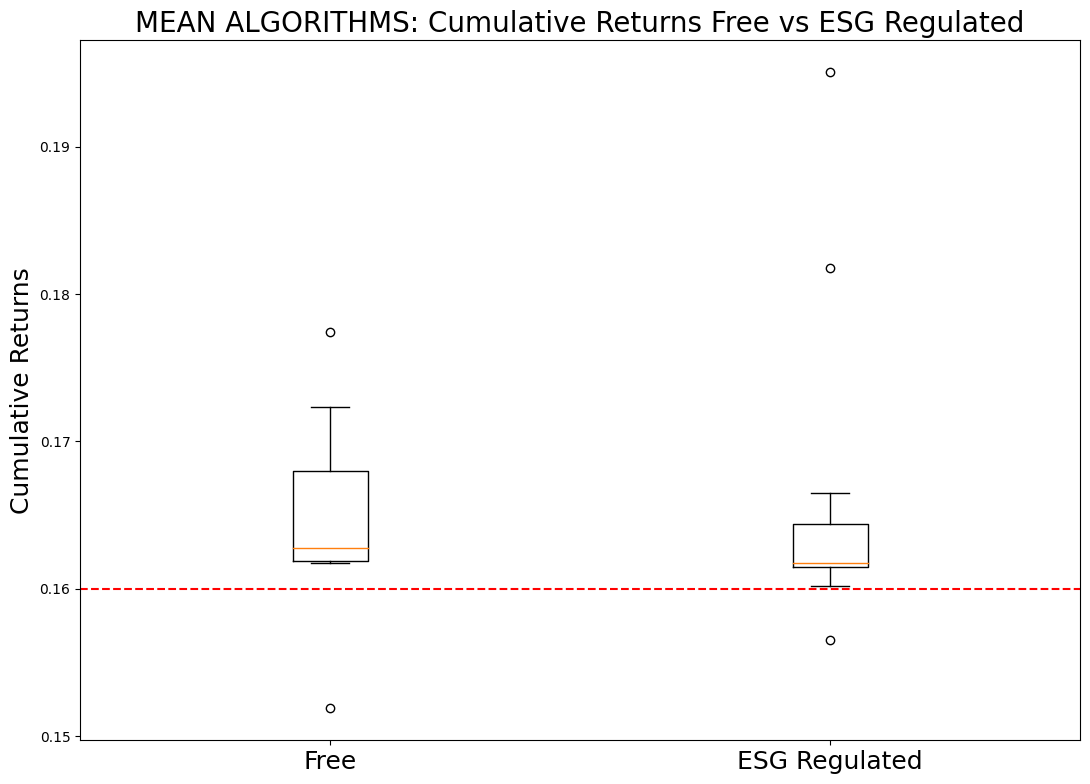}
    \caption{Mean cumulative returns of both agents regarding the performance of the A2C, PPO, DDPG, SAC and TD3 DRL algorithms.}
    \label{djia_boxplot}
\end{figure}

We have also conducted tests using the DDPG, SAC and TD3 algorithms on the DJIA, NASDAQ-100 and IBEX-35 indexes. To get rid from the bias that the particular algorithm introduces in the analysis we report mean metrics of the cumulative returns, Sharpe ratio and Annual Volatility in the following tables for the DJIA, NASDAQ-100 and IBEX-35 indexes. We can see in Table \ref{djia_norm} how, for all the algorithms, the main ratios of the ESG portfolio agent outperform the DRL agent operating in the free market. Differences are not very high, but we can not reject the null hypothesis that the performance of both agents is equal, which is enough statistical evidence that the DRL agent does not outperform the ESG DRL agent. We can also see in Figure \ref{djia_boxplot} how the maximum returns of the ESG agent are better than the ones of the free market agent, which suggests that, if conditions are stable and the hyper-parameters of the DRL agent are accurate the ESG DRL agent may outperform the free market agent. 

\begin{table}[h]
    \centering
    \begin{tabular}{|c|c|c|}
        \hline
        \textbf{Metric name} & \textbf{DRL A2C Agent ESG Regulated} & \textbf{DRL A2C Agent} \\
        \hline 
        Cumulative returns & $ \mathbf{0.1658 \pm 0.010}$ & $0.1649 \pm 0.006$ \\
        \hline
        Annual volatility & $\mathbf{0.1769 \pm 0.003}$ & $0.1792 \pm  0.006$ \\
        \hline
        Sharpe Ratio & $\mathbf{2.150 \pm 0.118}$ & $2.117 \pm  0.087$ \\
        \hline
    \end{tabular}
    \caption{Risk-performance metrics of the toy experiment in the DJIA index in the trading period according to their predictions with respect to its observable space for the experiments of all the DRL algorithms.}
    \label{djia_norm}
\end{table}

We now repeat the same operations with the IBEX-35 index, reporting the mean results of all the DRL algorithms in Table \ref{ibex_norm}. In this experiment, we show how the ESG DRL agent is outperformed by the free market agent. We hypothesize that different markets may produce different policies and that in this case being ESG is not as profitable as in the case of DJIA. However, differences are small and we did not have the ESG values of all the assets in this case, which might also explain the differences with respect to DJIA. 

\begin{table}[h]
    \centering
    \begin{tabular}{|c|c|c|}
        \hline
        \textbf{Metric name} & \textbf{DRL A2C Agent ESG Regulated} & \textbf{DRL A2C Agent} \\
        \hline 
        Cumulative returns & $0.086 \pm 0.017$ & $\mathbf{0.1078 \pm 0.012}$ \\
        \hline
        Annual volatility & $0.2693 \pm 0.019$ & $\mathbf{0.2595 \pm  0.006}$ \\
        \hline
        Sharpe Ratio & $0.875 \pm 0.176$ & $\mathbf{ 1.065 \pm  0.087}$ \\
        \hline
    \end{tabular}
    \caption{Risk-performance metrics of the toy experiment in the IBEX-35 index in the trading period according to their predictions with respect to its observable space for the experiments of all the DRL algorithms.}
    \label{ibex_norm}
\end{table}

\begin{figure}[h]
    \centering
    \includegraphics[width = 0.7\textwidth]{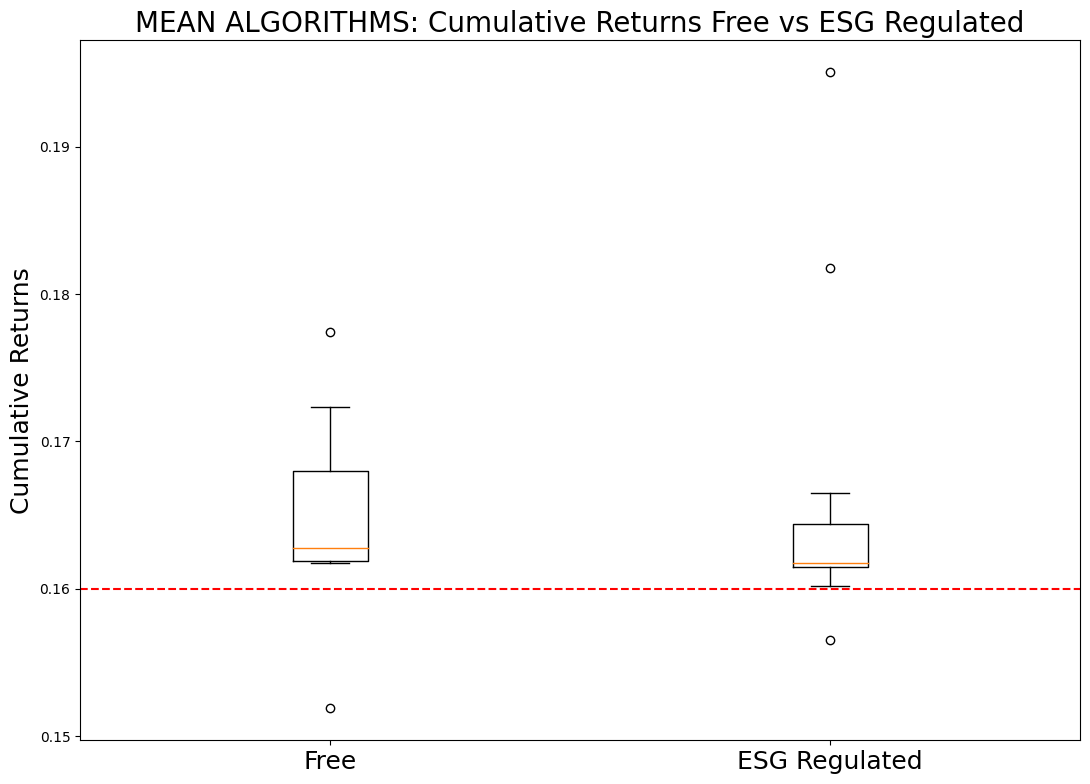}
    \caption{Mean cumulative returns of both agents regarding the performance of the A2C, PPO, DDPG, SAC and TD3 DRL algorithms.}
    \label{nasdaq_boxplot}
\end{figure}

Finally, we do the same experiments for the NASDAQ-100 index, to add more empirical evidence to the claim that configuring a ESG policy does not harm the performance of a DRL agent. We report the mean results of all the DRL algorithms in Table \ref{nasdaq_norm}. We can see how in this market, a better performance is achieved by the ESG policy but the risk is slightly lower in the case of the free market policy. However, although the mean of the Sharpe ratio is bigger in the case of the free market agent, we can observe in Figure \ref{nasdaq_boxplot} how the maximum value is actually bigger in the case of the ESG market, displaying more variability in its behaviour and being its mean harmed by one low Sharpe value. We can see how both agents generally outperform the baseline, which is the index strategy.    

\begin{table}[h]
    \centering
    \begin{tabular}{|c|c|c|}
        \hline
        \textbf{Metric name} & \textbf{DRL A2C Agent ESG Regulated} & \textbf{DRL A2C Agent} \\
        \hline 
        Cumulative returns & $\mathbf{0.4580 \pm 0.038}$ & $0.4567 \pm 0.0157$ \\
        \hline
        Annual volatility & $0.1666 \pm 0.0020$ & $\mathbf{0.1651 \pm  0.0016}$ \\
        \hline
        Sharpe Ratio & $1.7753 \pm 0.1193$ & $\mathbf{ 1.7865 \pm  0.0400}$ \\
        \hline
    \end{tabular}
    \caption{Risk-performance metrics of the toy experiment in the NASDAQ-100 index in the trading period according to their predictions with respect to its observable space for the experiments of all the DRL algorithms.}
    \label{nasdaq_norm}
\end{table}

Summing up, we have provided lots of empirical evidence using different algorithms and indexes that show how we can not reject the null hypothesis that both policies deliver the same results, in statistical terms. This implies that the statement that ESG policies deliver worse results for an investor using robotraders trained by DRL algorithms cannot be proved, ergo if both policies are similar, investing in a ESG policy has the benefit of promoting companies that care about ESG metrics. 

\section{Conclusions and further work}

In this work, we have studied deep reinforcement learning (DRL) for ESG financial portfolio management. We conducted several tests on the Dow Jones Industrial Average (DJIA), IBEX-35, and NASDAQ-100. The results have yielded significant insights, underlining the potential of DRL agents in ESG-driven portfolio management. Our research implemented a dual market approach for comparative analysis. We assessed the performance of DRL agents in both unregulated markets and in markets where financial incentives and penalties were imposed based on the ESG score of the portfolio in relation to the weighted mean ESG score of the index. The performance was evaluated using multiple metrics such as Sharpe ratio or the cumulative return, among others. Remarkably, we found that including ESG information into the state space of the agents did not adversely affect the performance of the portfolio. Moreover, the devised DRL agent demonstrated promising results in terms of adhering to an ESG policy for portfolio management. 

For further work, we propose the use of safe reinforcement learning methodologies \cite{garcia2015comprehensive}, customized for ESG-oriented investment decision-making. Secondly, we aim to leverage multi-objective Bayesian optimization for the tuning of hyper-parameters. This approach can help to simultaneously optimize a risk-performance metric and an ESG metric, thereby balancing financial and sustainability aspects more effectively \cite{garrido2021advanced}. Finally, we believe exploring causal reinforcement learning \cite{zhu2019causal} could offer intriguing insights. If ESG information indeed has causal relationships with risk-performance measures, it could provide a deeper understanding of how ESG factors influence investment outcomes.
\bibliography{main}
\bibliographystyle{acm}

\end{document}